\documentstyle[12pt]{article}

\begin{document}

\begin{flushright}
IJS-TP-94/25
\end{flushright}
\vskip 1cm
\centerline{\bf EVIDENCE ON $qq\bar q\bar q$ HADRON SPECTRUM }
\baselineskip=22pt
\centerline{\bf FROM $\gamma\gamma\to vector\,meson\,vector\, meson$
REACTIONS ?\footnote{Talk presented at the {\it International
Conference on Quark Confinement and the Hadron Spectrum},
Como, 20.-24.6.1994. To be published in the Proceedings.}}

\vspace{0.8cm}
\centerline{\rm MITJA ROSINA}
\baselineskip=13pt

\centerline{\it Faculty of Natural Sciences and Technology,
and Jo\v zef Stefan Institute,}
\baselineskip=12pt
\centerline{\it University of Ljubljana, Jadranska 19,
P.O.B.64, 61111 Ljubljana, Slovenia}

\vspace{0.3cm}
\centerline{\rm and}
\vspace{0.3cm}
\centerline{\rm BORUT BAJC}
\baselineskip=13pt
\centerline{\it Jo\v zef Stefan Institute, Jamova 39, p.p. 100,
61111 Ljubljana, Slovenia}
\vspace{0.9cm}
\abstract{
The $\gamma\gamma\to\rho^o\rho^o\to 4 \pi$ reaction shows a
broad "resonance"at 1.5 GeV with no counterpart in the
$\rho^+\rho^-$ channel. This $(J^P,J_z)=(2^+,2),
\;I=0 \, \hbox{and} \, 2$ resonance is considered
as a candidate for a $qq\bar q\bar q$ state.
We show, however, that it can also be explained
by potential scattering of $\rho^o\rho^o$ via the $\sigma-$ exchange.}

\rm\baselineskip=14pt
\vskip 0.6cm
{\bf 1. The Motivation}
\vskip 0.4cm
After many successes of quark models
to describe a single meson or baryon,
their predictive power in the two-hadron
sector remains questionable.
A fair description of the
nucleon-nucleon interaction was accompanied
with the prediction of a rich dibaryon spectrum which has never
been observed. It is
interesting to see how different quark models
perform in the two-meson sector,
as compared to mesonic models. We are
also motivated by our experimental colleagues in Ljubljana
(ARGUS coll.\cite{ARGUSONE}$, $\cite{ARGUSTWO}).

\vskip 0.6cm
{\bf 2. The Experimental "Puzzle"}
\vskip 0.4cm
The $\gamma\gamma\to\rho\rho$ reaction
\cite{ARGUSONE}$, $\cite{ARGUSTWO}
has a large and broad peak near
the nominal threshold in the
$\rho^o\rho^o , (2^+2)$ channel
and a much smaller cross section
in other $\rho^o\rho^o $
channels as well as in the
$\rho^+\rho^-$ channel.
Since so far the quark models \cite{ACHASOV}
covered more features simultaneously
than the effective mesonic models \cite{MOUSSALLAM},
the opinion prevails that
explicit quark degrees of freedom
(a $qq\bar{q}\bar{q}$ molecule)
are essential. Here
we challenge this view.

\vskip 0.6cm
{\bf 3. The Hypothesis}
\vskip 0.4cm
The peak near the nominal threshold and its strong spin dependence
are reminiscent of the p-n scattering at low energy.
There the S=0 channel has a very high cross section while
in the S=1 chanel it is much lower, because in the S=0
state the potential well contains almost exactly 1/4 wavelength
of the relative wavefunction.

We hypothesize that a similar trick
of nature is played also in $\rho\rho$ scattering.
We assume that both photons are converted
in $\rho$ via vector dominance, and both
$\rho$ then interact by a scalar isoscalar
potential which is just strong
enough to (almost) bind them for one spin
orientation and fails to do so
for the others. Also, an isoscalar potential
cannot exchange charges and does not lead to
the $\rho^+\rho^-$ final state.

\vskip 0.6cm
{\bf 4. A Toy Potential Model}
\vskip 0.4cm
To illustrate this idea, we choose a Yukawa-type
$\sigma$-exchange potential with the same parameters as in the
Bonn potential for the nucleon-nucleon system \cite{MACHLEIDT}
($g^2/4\pi=7.07,\,m_{\sigma}=0.55$ GeV), but
multiplying it with a factor $({2\over 3})^2$
(two quarks rather than three at each vertex).
Solving the nonrelativistic Schr\"odinger equation
we showed that with a very plausible
hard core ($r_c=0.17$ fm) a weakly bound
or antibound state at $E\sim 0$ can be obtained.
If the potential is assumed
to be slightly spin dependent one
spin channel will have a state
close to zero and other spin channels
will miss it. This demonstrates
that the proposed hypothesis can be realized.

\vskip 0.6cm
{\bf 5. A Relativistic Potential Model}
\vskip 0.4cm
Our starting Lagrangian
is gauge invariant and respects vector dominance:

\begin{eqnarray}
\label{LAGRANGIAN}
{\cal L}=&-&{1\over 4}(\partial_\mu B_\nu-\partial_\nu B_\mu)^2
-{1\over 4}(\partial_\mu\rho_\nu-\partial_\nu\rho_\mu)^2
+{1\over 2}m_\rho^2\rho_\mu^2
+{1\over 2}(\partial_\mu\sigma)^2-{1\over 2}m_s^2\sigma^2\nonumber\\
&-&{1\over 2}m_\rho^2[\rho_\mu^2-(\rho_\mu-
{e\over g}B_\mu)^2]
+{g_s\over 2}\sigma(\partial_\mu\rho_\nu-
\partial_\nu\rho_\mu)^2\;.
\end{eqnarray}

The amplitude was calculated with the Bethe-Salpeter
equation in the leading order of $e/g$
with the $\sigma$ exchange kernel
given from Eq. (\ref{LAGRANGIAN}).
After the Blankenbecler-Sugar reduction \cite{BBS}
and the partial wave expansion \cite{JW}
the integral
equations for different channels were solved
with the matrix inversion method \cite{BJKHT},
taking $16$ points for
the magnitude of the relative three-momentum from
$0$ to the cutoff $\Lambda$.
Assuming that the effect of different off-shellness
of the final $\rho$ is small,
we got the cross section by weighing the
amplitude squared with Breit-Wigner factors.

We choose for the vector dominance factor $e/g=0.036$,
which is consistent with the value obtained within a
larger model \cite{GGLN} from the $\rho$ and $a_1(1260)$
decay widths. The mass in the two-body propagator
was chosen $m=0.692$ GeV instead of $m=m_\rho=0.77$ GeV,
since the $\rho$ meson is broad.
The results proved to be relatively insensitive
to the choice of the $\sigma$ mass, so that we will
report only the case $m_s=0.5$ GeV.

The results are shown in fig. 1,
for different choices of
the paramaters $\Lambda$ and $g_s$ together
with the experimental values. A combination of the
two parameters in a reasonable range exists,
which reproduces quite well the experimental results.

Some models \cite{ACHASOV} predict, that
the enhancement
in the $(2^+,2)$ channel is a universal feature of
$\gamma\gamma\to 2\, vector\, mesons$.
Experimentally, there are too few events to judge.
Due to the sensitivity to potential parameters in our models
we predict that such enhancement is very difficult to appear
in more than one case.

Work supported by the Ministry of Science and
Technology of Slovenia.

\vskip 1cm

\centerline{FIGURES}

\vskip 0.5cm
\noindent
Figure 1: Cross sections for various channels in
the relativistic potential model with scalar
exchange: --- ($\Lambda=2.0$ GeV, $g_s=12.0$
GeV$^{-1}$); $\cdots$ ($\Lambda=2.2$ GeV, $g_s=10.4$
GeV$^{-1}$); - - ($\Lambda=2.4$ GeV, $g_s=9.13$
GeV$^{-1}$); experiments: $\bullet$ ARGUS \cite{ARGUSONE},
$\circ$ ARGUS \cite{ARGUSTWO}.

\end{document}